\documentstyle[preprint,epsfig,aps]{revtex}

\begin{document}
\draft
\title{\bf Topological Excitations in Doped Spin Ladders}
\author{Yujoung Bai}
\address{National Creative Research Initiative Center for Superconductivity, 
Department of Physics,
Pohang University of Science and Technology, Pohang, Kyungbuk  790-784, KOREA}

\maketitle
\begin{abstract}

 We study low-energy magnetic excitations of doped spin-ladders,  based on an effective Hamiltonian describing  interactions of mobile spin and 
 background spins.   The helicity moduli against fluctuations    are calculated in terms of the doping 
 concentration, the leg number and interaction strengths in the ladder.   
  The doping ranges for spiral phase are 
 computed and its relation to spin gap structure is discussed. 
 The spiral order  can enhance existing antiferromagnetic order at 
 certain doping range or reduce it at very low density of holes. 
 For odd-legged ladders,  formation  of spin gap at certain narrow 
 doping range  becomes possible via coupling of spirals and background spins. 
 \end{abstract}
 \vspace*{1cm}

	There have been numerous studies on long-range magnetic order in two-dimensional quantum systems, some of which employed mapping to (2+1)D non-linear sigma model(NLsM)\cite{chakravarty} for the long-wavelength, low-energy states. Adding the effects of doped holes to 2D antiferromagnets,
several works were done of the in-plane spiral modes by mean-field calculation \cite{auerbach}\cite{kane}\cite{arrigoni}\cite{mori}. A couple  of studies 
  included the out-of-plane spirals generated in 2D systems, starting with t-J models\cite{hamada} which incorporated the effects of  mobile spin in magnetic background\cite{shraiman}.
  By  spiral modes, we mean  vortex rotations formed by background spin phases or long-wavelength excitations  arising
   from fluctuations of staggered magnetization. In lower dimensions, some studies have been done for low-energy excitations 
   in $\it{undoped}$ spin ladders\cite{senechal} \cite{Dell'Aringa}\cite{normand}.
   The ground state of even-legged ladders is known to be spin liquid(singlets), given strong to intermediate coupling -- the exchange ratio being
    $J_{\perp}\ge J_{\parallel}$. The odd-legged ladders have
     even-parity channels and odd-parity channel, due to the reflection symmetry. 
    The mean-field theory(MFT) and numerical studies agree that the  even-channel makes the  spin liquid while the odd-channel makes the Luttinger liquid, before doping. After started doping, the hole spins tend to go  into the even-channel, thus keeping the gapped mode just like even-ladders, much more likely than into the odd-channel. 
   It has been mostly accepted that the ``topological term'' in the NLsM describing a $\it{undoped}$ ladder is classified by the parity of leg numbers($n_{l}$).
   The term is shown to be zero for undoped even-legged  ladders and $2\pi S\hspace{0.1cm}n_{l}$
    for odd-legged  ladders\cite{sierra}. As for $\it{doped}$ ones, there have been a couple of results which showed clear departure from this type of distinction, by
	DMRG calculations\cite{white} and by Lanczos calculations along with  MFT\cite{rice}.
 In this paper, we focus on the low-energy spin excitations  in doped ladders by analytical calculations, discuusing the role of the topological term for doped ladders. Doped spin ladders consist of  background spins  and empty sites with residual spin. The effective interaction between residual spins are antiferromagnetic, if they belong to different sublattices of any bipartite ladder sites\cite{sigrist}.
 New magnetic excitations arise from  coupling of collective polarization and local magnetization, as the effective spin of the hole  moves around the ladder sites. In the continuum limit, the distorted spin background induces non-zero ``magnetic" flux due to phase fluctuation. 
   To see the spin response, the helicity modulus is calculated as an
   explicit function of doping rate, leg number, temperature and interaction strengths in spin ladders.
  By minimizing the total energy of the system, we find  the range of  hole concentration which favor the spiral mode, from the helicity modulus.
  Our results of the doping range for gapped mode in odd-ladders agree well with those existing numerical studies\cite{rice}\cite{white}, within reasonable range of interaction coupling in our model.
 Also, we determine how spiral modes enhances/suppresses  
    AF order and the range for these effects.
	
   We  begin with an effective Hamiltonian, based on the one made by Shraiman and Siggia for 2D magnets\cite{shraiman}. At half-filling, the low energy state has the commensurate Neel order. For the long-wavelength staggered spin state,
 the  Neel vector is twisted,
 by $\alpha{y\over L}$ around z-axis at the end of  the ladder length  $L$  with respect to the first $\widehat{n}$ and by $\nu{x\over (n_{l}-1)a_0}$ at the end of the ladder with $n_{l}$-leggs (with the lattice spacing $a_{0}$). When the mobile spins are introduced upon doping, the Neel vector is allowed to rotate around y-axis as well, to derive all modes generated by collective polarization and phase fluctuation. $\widehat{n}=(\sin \theta \cos\phi, \sin \theta\sin\phi, \cos\theta)$. This is equivalent to twisting the spin quantization axis from site to site
 via appropriate transformation as done in some existing works\cite{arrigoni}\cite{kane}\cite{hamada}. The effective Hamiltonian is 
\begin{equation}
H=H_{NLsM} + {1\over 2}\left({{\dot{y}\alpha}\over L}\right)^2+
 {1\over 2}\left[{{\dot{x}\nu}\over (n_{l}-1)a_0}\right]^2- 
\widetilde{g}\sum_{a,q} \widehat{P}_{a}(q)\cdot \vec{j}_{a}(q)
- g\prime\sum_{a,k}\cos k \bar{\Psi}_{k-{2\over q}}\widehat{\tau} 
\Psi_{k+{2\over q}}
\cdot\widehat{m}(q)
\end{equation} 
where the magnetization current $\vec{j}_a(q)= \widehat{n}\times
\nabla _{a} \widehat{n}$
 and the dipole  polarization $\widehat{P}_{a}(q)=\sum_k \sin k_a \bar{\Psi}_{k-{2\over q}} \hspace{0.1cm}\widehat{\tau}\hspace{0.1cm}\Psi_{k+{2\over q}}$.
  The second and the third term is the rotational kinetic energy of the unimodular $\widehat{n}$.  The subscript $a$ stands for in-plane(IP) mode and out-of-plane(OP) spirals.
 The wavevector of the mobile spin  is related to the spiral wavenumber $Q_a$  as $<\bar{\Psi}_{k-{2\over q}}\hspace{0.1cm}\Psi_{k+{2\over q}}>\approx Q_a \sin k_a$. Identifying the components
 of incommensurability wavevector $\vec{Q}=(Q_x, Q_y)$  with the angular terms  in the effective interaction term in $H_{eff}$, 
 the spiral wavevector is $\vec{Q}\rightarrow  -\sin\theta\cos\theta(\cos\phi+\sin\phi)]\equiv
 \cos(Q_x/2)-\cos(Q_y/2)$.
   The spiral wavevector  has a symmetry similar to 
 a d-wave. The effective  propagation in real space is along the rung 
 direction of the ladder, having arisen from
    triplet formation along the ladder direction. This mode is equivalent to the
  transverse mode\cite{arrigoni} or to the stripe phase\cite{mori} 
  as termed for 2D systems\cite{auerbach}. 
  The doping range for the spiral order to occur is computed from the 
  helicity modulus against IP twisting. 
  The  spiral  wavenumber $Q$ is proportional to the doping rate\cite{kane}\cite{hamada}.
  For the purpose of calculating the helicity modulus, we use $Q\approx \delta {{t_\perp}\over {J_\perp}}$ for IP mode and $Q\approx \delta$ for OP spirals. 
 For IP mode, the effective Hamiltoinian density is given
\begin{displaymath}
H_{IP}=(\partial_a\psi)^2 + {1\over 2}\left({{\dot{y}\alpha}\over L}\right)^2+
 {1\over 2}\left({{\dot{x}\nu}\over (n_{l}-1)a_0}\right)^2 -
\end{displaymath}
\begin{equation}
 2\widetilde{g}\delta{{t_\perp}\over {J_\perp}}\{\cos\phi(\cos\phi+\sin\phi)\}\sin^2 k 
\pm 2g\prime\delta{{t_\perp}\over {J_\perp}}\cos\phi\sin k\cos k
\end{equation}
The free energy per unit area of the ladder plane is 
\begin{displaymath}
F=-{1\over \beta}\ln Tr \exp[ -\beta H_{eff}]
=-{1\over \beta}\ln \left[{2\pi(n_l-2)\over a_{0}^2(n_l-1)}\right]+
\beta\left[{1\over 2}\left({{\dot{y}\alpha}\over L}\right)^2+
 {1\over 2}\left({{\dot{x}\nu}\over (n_{l}-1)a_0}\right)^2\right]
\end{displaymath}
\begin{equation}
+{1\over 2\beta}\ln[2{\beta}^2\{1+A\cos\phi(\cos\phi+\sin\phi)\}]
-{\beta B^2 \cos^2\phi\over 4\{1+A\cos\phi(\cos\phi+\sin\phi)\}}
\end{equation} 
where $A=-2\widetilde{g}\delta{{t_\perp}\over {J_\perp}}$, $B=-2g\prime\delta{{t_\perp}\over {J_\perp}}$ and
 $\beta=1/ {k_B}T$. The subscript $\perp$ refers to those along the rung direction while $\parallel$ is for those in the chain direction.
 The spin wave velocity for doped ladders is derived from 
 mapping the unit ladder cell (repeated regularly as defined by the density of empty sites) to NLsM. 
  Taking only the slow component of the Neel vector and for  the coherence length 
$\gg$ the ladder width($a_{0} n_l$), 
   the coupling  now includes  the doping effect
 $g= [(1-\delta)n_{l}S]^{-1}
 \left(1+{J_{\perp}\over 2J_{\parallel}}\right)^{1\over 2}$.
      The spin wave velocity is collected from the prefactor,
 $v_s=2SJ_{\parallel}\left({{1-2\delta}\over {1-\delta}}\right)
 \left(1+{J_{\perp}\over 2J_{\parallel}}\right)^{1\over 2}$.
As for 2D systems without any hedgehogs, it is known that the topological term $-{\theta\over 4\pi}\int{(\partial_{a}m)^{2}}d^{2}r$ is zero, regardless the spin number or the configurations of lattice sites\cite{haldane}\cite{fradkin}\cite{ioffe}.  We discuss the different meaning of this term in doped ladders later.
 The coupling in the effective interaction($V_{g\prime}$) is found from
relating the hole density to the collective polarization $<P_{g\prime}>$ and the incommensurability wavenumber $Q$ of the spiral mode. From 
 $<V_{g\prime}>\simeq 2g\prime\hspace{0.1cm} Q\ln[2(n_l-1)]<P>\simeq g\prime\hspace{0.1cm}Q(n_l-1)L/ 16\pi^2 a_{0}$, 
  we estimate the coupling 
$g\prime\approx {J_{\perp}Q/ <P>}\approx {32a_{0}\pi^2\ln[2(n_l-1)]\delta J_{\perp}/ L(n_{l}-1)}$, 
 where $L$ is the length of unit ladder cell, which decreases as hole density  increases.  The other coupling $\widetilde{g}$ can be estimated from renormalized coupling($g_\sigma$) of the NLsM;
  $1/\widetilde{g}=1/g_{\sigma}$+Tr$[1/\{k^2 + V(\widetilde{g})\}]$. For instance, given the most realistic interaction strengths in cuprate ladders ${t_\perp\over J_\perp}=5$,  $\widetilde{g}= 1$ with $\delta=0.1$, while $\delta\ll 1$ for  much greater   $\widetilde{g}$.
   The helicity modulus(HM) against the in-plane fluctuations is computed 
\begin{displaymath}
 \Re_{IP} = {v_s\over 2}\left|{\partial^2 F\over \partial{\phi^2}}
 \right|_{\phi=0}
\end{displaymath}
\begin{equation}
 =\left({{1-2\delta}\over {1-\delta}}\right)
{(\widetilde{g}\delta t_{\perp})^2\over {\beta J_{\perp}(1-
2\widetilde{g}\delta{{t_\perp}\over {J_\perp}})^2}}
\left(1+{J_{\perp}\over 2J_{\parallel}}\right)^{1\over 2}
\left[1-{(128\pi^2\beta\ln[2(n_l-1)]a_{0}\delta^2 t_{\perp})^2\over {(n_l-1)^2(2 \widetilde{g}\delta{t_\perp\over J_\perp}-1)}}\right]
\end{equation}
This is  an explicit function of the doping concentration $\delta$, leg number $n_l$, temperature and the interaction strengths in the ladder $t_\perp$, $J_{\perp}$ and $J_{\parallel}$.
 Fig. 2  is a surface plot of the IP HM versus doping rate
and temperature.  The common features of HM for all ladders are (i) that the HM diverges
at  certain doping rate $\delta=1/(2\widetilde{g}{t_\perp\over J_\perp})$, 
(ii) that the  sudden jump is followed by sharp decrease in HM to
negative values, (iii) that the flat plateau is over wide range of doping
and temperature, (iv) that another range  for shrap increase of HM is at 
$0.8\leq \delta \leq 0.9$. 
  The spin liquid ground state acquires excited triplets
 upon doping. When the hole density reaches a certain 
level, intrachain pair formation becomes substantial as well,
 which makes the whole system very stiff. The diverging modulus followed by 
 drastic decrease as well as flattening
  over wide range,  indicate  presence of spin gap, since the 
  HM(stiffness) goes  as the inverse of spin susceptibility.
  In the region of  negative HM, the system releases spiral modes
  to lower the energy. Though  HM itself is a local response, the spiral  phase (bulkwise) can be formed in the regime of negative HM. As doping increases further, there is another sharp increase in HM at very low temperatures, due to  coupling of spirals and intrachain 
   spins.  The spiral modes can put a pair of spins 
 into a level $\it{in}$
 the gap.  Thus, the short-range AF order of spin liquids can be lost
  at certain doping regime.  In even- ladders with existing  spin gaps,
  spiral modes can reduce the spin gap sizes. 
 At higher  $\widetilde{g}$ which is also physically feasible, the spiral
phase is absent at low doping, though the other window at much higher doping still remains. In Fig. 2 for 3-leg ladder, the possibility of generating 
  spin gap in odd-ladders exists  around $\delta=0.1$ with $\widetilde{g}$=1. For more attractive $\widetilde{g}\geq 1$ which is very likely in physical ladders, the spiral mode and gap formation occur in smaller  doping regime of $\delta < 0.1$,  at  temperatures $ T \leq 5 K$. 
  The region where the HM becomes negative is commonly  ${J_\perp\over 2\widetilde{g}t_\perp}< \delta < 0.5$ at  $T\le 5 K$,  producing spiral phases for all ladders. The characteristics are repeated for ladders with any number of leggs, though precise values are a bit different, due to the leg-number dependence in $g$.   As leg-number increases, the doping regime for IP spiral mode is reduced. Since the even-parity channels in odd-ladders are dominant over the odd-parity channels, the added(effective) spins tend to be in the even-channels.  Given the attractive potentials provided
	  by $V_{g}$'s in our model, binding of the spins in the odd-channel and the spirals(solitons) yield an energy gap. This gives rise to a gapped phase  at $\delta\ge \delta_{spiral}$ in odd-ladders, with all channels participating. Our  result of doping rate for the gapped mode in odd-ladders ($\delta_{spiral}=0.1$ for $\widetilde{g}= 1$) is close to  the result by Rice et. al.
 \cite{rice} who found
  the $\delta_{c}\ge 0.13$. If the coupling  $\widetilde{g}> 1 $ in physical ladders, which gives our $\delta_{spiral}< 0.1$, the
  DMRG result of $\delta_{c}\ge 0.06$ by White et. al\cite{white} is  comparable as well.
    In our picture, there is interplay between effective interaction strengths and hole concentration, giving the coupling $g$ vary consequently. Existing interpretation of gap generation in odd-ladders by White $\em{et. al.}$ uses the idea of domain walls formed by  competition between the  kinetic energy and the exchange energy. In our picture, domain walls can be interpreted as areas of the spiral  vortices  with different spiral wavenumber or  different chiralities.  Alternative understanding has been given with proximity effect to the gapped mode of the even-channel, which  enhances the pairing within the odd-channel\cite{rice}\cite{emery}. 
	  
  For out-of-plane(OP) mode, the effective Hamiltonian for OP spiral is $H_{OP}=(\partial_a\psi)^2\pm 2g\prime\delta\sin k\cos k\cos\theta$.
 The free energy per unit area along the $\widehat{z}$ is
$F={-1\over \beta}\left[\ln[{2\pi\over \beta}]+2(g\prime\cos\theta\delta\beta)^2\right]$.
The helicity modulus(HM) against the out-of-plane twisting is
 \begin{equation}
\Re_{OP}= {v_s\over 2}\left|{\partial^2 F\over \partial\theta^2}
 \right|_{\theta=0}=
 -2\beta\left({32{\pi^2}\ln[2(n_l-1)]\over (n_l-1)}\right )^2\left({1-2\delta\over 1-\delta}\right)\delta^4
J_{\perp}^3\left(1+{J_\perp\over 2J_\parallel}\right)^{1\over 2}
\end{equation}
 For a exchange ratio of $J_\perp= 2J_\parallel$, the result shows formation of OP spiral phase over wide doping range 
 $ 0 <\delta < 0.5$ for ladders with any number of $n_l > 1$. In all ladders, there are 
 sharp increases in OP HM at $\delta\ge 0.5$ for $T < 15 K$,  making
   OP spirals robust. 
 At very high hole density  beyond this $\delta$, the peaked modulus indicates forming  bound states of
	OP spirals  with opposite chirality(soliton-antisoliton pairs).  The spin quantum number would be   integers for this new type of  condensate. The HM(stiffness) decreases as temperature increases at high doping regime for all ladders.  
  From  minimizing the Hamiltonian $H_{eff}$ with respect to a mean-field $Q$\cite{shraiman},
  the self-consistency condition gives
  $Q={\widetilde{g}\over \Re}\sum_k\sin{k}\hspace{0.1cm}(n^{+}+n^{-})
  \simeq {2\widetilde{g}\beta\over J}\sum_k\sin{k}\hspace{0.1cm}(\mu-\varepsilon_{k})$.
  The Fermi distribution function 
      $n^{\pm}=\exp[\beta(\varepsilon_{k}\mp \widetilde{g}\hspace{0.1cm}Q\sin{k}-\mu)]+1$
   	gives $n^{+}-n^{-}\approx 2\beta\widetilde{g}\hspace{0.1cm}Q\sin{k}$
  	 at low $T$.
 Then the mean effective interaction terms are
 $<V_{\widetilde{g}}> + <V_{g\prime}> 
 \simeq-{\widetilde{g}^2\over \Re}\sum_{k}\sin^{2}k\hspace{0.1cm}(2\beta\widetilde{g}\hspace{0.1cm}Q\sin{k})^2 + {1\over 2}\Re Q^2$.  The $<H_{MF}>$ becomes positive when the helicity modulus $\Re$ becomes negative, which makes the system unstable. So, the system releases the torsional momenta. That is, the spiral phase with uniform wavenumber throughout the system is favoured when 
  $\Re < 0$. From  $\Re_{IP}$, we find the doping range for spiral phase 
${J_\perp\over 2\widetilde{g}t_\perp}\le\delta< {1\over 2}$, which agree with the graphical result. 
 The spin coherence length in the ladder plane is computed when the HM is zero,
  \begin{equation}
  \xi\simeq{128a_{0}\pi\ln[2(n_{l}-1)]\delta\beta J_{\perp}
   \over (n_{l}-1)[2\delta\widetilde{g}t_{\perp}/J_{\perp}-1]^{1/2}}
  \end{equation}
 When the incommensurate spin state is stable in the regime of positive
  HM, it is possible to gather the (hole) spins along valleys of the spiral wavevector. This may result in  phase separation of hole-rich region and no-hole region\cite{arrigoni}\cite{orignac}, adjacent to the spiral phase.
 As for spin ladders with even number of leggs, the upper limit of doping for Luttinger liquid phase is  $\sim J/2t\widetilde{g}$. The lower
   limit would be around $J/4t\widetilde{g}$. 
   In our results, there is no transition like Kosterlitz-Thouless(KT) transition in 2D\cite{kosterlitz}, since  there is no strong size dependence and no plateau in the stiffness around any particular temperatures. Though there are size dependent terms in the expression of the HM, these are rather weak.  Calculated spin coherence suggests that the gap function has a different form from this, due to the doping effect.
  As far as the helicity modulus, ladders with even/odd-number of leggs do not have differing signs, since the factor $\ln[2(n_l-1)]$ does not distinguish 
   the parity in total leg number $n_l$. This is the difference from the result on 1D spin chains which distinguish the parity, but analogous to 2D systems which do not differentiate the spin quantum number being integer or half-odd integer. For ladders with finite number of leggs, the boundary condition   in the spin wavefunction is given 
$\Psi_{l}(n+1)={\pm}\exp[-i\phi]\Psi_{1}(1)$, where $l$ is the leg index and $n$ is the site number in the block of the system\cite{loss}.
 The  meaningful parity  in ladders is that of (leg number plus site number), instead of the leg munber alone. The ``+'' or ``-'' sign goes as the chirality associated with the winding orientations of spirals,  having arisen from the 
   presence of two sublattices(of up or down-spin).
So, the ground state in $\it{odd-ladders}$ is degenerated for spiral modes with different chiralities at different ladder regions, thus   satisfying Lieb-Mattis-Shulz theorem. This means that gapped mode is possible in odd-ladders. This is
 the same result as found by previous MFT study on odd-ladders\cite{sigrist}.

  So far, we restricted the spiral winding number ($w$) to 1, to ensure the spinor wavefunction to be  periodic (instead of anti-periodic). 
 Corresponding staggered spin flux(chirality) in continuum limit is $2\pi w$. In terms of the Neel vector at bipartite ladder sites, the winding number is
$=[\widehat{n}_{1}\cdot\widehat{n}_{2}\times\widehat{n}_{3}+
 \widehat{n}_{1}\cdot\widehat{n}_{3}\times\widehat{n}_{4}-
 \widehat{n}_{1}\cdot\widehat{n}_{2}\times\widehat{n}_{4}-
 \widehat{n}_{2}\cdot\widehat{n}_{3}\times\widehat{n}_{4}]$.
 If the systems were undoped ladders described by NLsM, the $w$ would be equal to the topological term.
 This gives the solid angle subtended by the Neel vectors or the Berry phase acquired by the Neel vector while traversing around the path in spin space
 \cite{baskaran}\cite{lee}.
  Allowing the spiral modes, the winding number need not be integers. This implies that   the  topological term in $\it{doped}$ ladders does not simply
  determine their classes by the leg numbers.
   The induced field acting as twisted boundary would affect the magnetic ordering in ladders. To see whether spiral modes compete with antiferromagnetism in $\it{ladders}$ or not, we rewrite the Neel vector including the winding number.
$\widehat{n}=(\sin \theta \cos (w\phi), \sin \theta\sin (w\phi), \cos\theta)$.
 Then, the effective interaction term  is  
  $V_{\widetilde{g}}\sim -\pi^2 w\widetilde{g}\delta$, coming from the gradient of $\widehat{n}$.
 This can provide attractive potential\cite{john} between the polarised spin and the spin current if the $w$ is positive. To maintain the attractive potential  between the spin texture and the spiral(soliton), $w$ is chosen to be of the chirality ``+''. This would result in destroying the AF order of background spins. However, there is another interaction term $V_{g\prime}$ in our effective Hamiltonian, which does not depend on the winding number. This  contribution can be greater in length scales of a few lattice constant, though it decreases as the doping decreases and/or the leg number increases. In cases of very low hole density (a few percent) and  many number of leggs, it is possible that the $V_{\widetilde{g}}$ wins to reduce the AF order.
  A known result by Arrigoni et. al. on spiral order in 2D Hubbard model   shows that the transverse  mode  $\it{enhances}$ AF order. The other spiral mode can enhance ferromagnetic order of the background. Another work by Mori et. al. on 2D  $t-J$ model also show two similar spiral phases.
  As for doped $\it{ladders}$ which started with commensurate Neel order at half-filling, our result show that the whether spiral modes  reduce or enhance the AF order depends directly on the hole density. Only at very low doping regime (a few percent), spiral order suppresses antiferromagnetism. It is  due to that the  effective interactions $V_{g\prime}$ and $V_{\widetilde{g}}$   compete in some doping regime. 
  
 In summary, due to the hole motion distorting the spin texture in doped 
 ladders, phase fluctuations of the staggered magnetization yields new type of excitations. In continuum limit, 
 long-range spiral modes arise and give the helicity modulus vary depending on the
  doping rate, temperature, leg number and interaction strengths.
 The spiral modes can put magnons
 into  sublevel(s) in the gap. The spin gap is reduced  in even-ladders while it can be generated in odd-ladders, depending on hole density.  Since the even-parity channels in odd-ladders are dominant over the odd-parity channels, the added spins tend to be in the even-channels. 
  Thus, in the begining of doping, the even-channels give the spin gap as
   spin liquids of undoped ladders.
  Given the attractive potentials  provided
 	   in our model, the spin current in the odd-channels tend to make bound states with the spirals in the even-channel. The enhanced pairing brings out the gapped Luther-Emery phase at $\delta\ge \delta_{spiral}$ in odd-ladders. 
 Our results of the doping range for gapped mode in odd-ladders agree well with those existing numerical studies.
 Antiferromagnetism in ladders can be enhanced over substantial doping range via having spiral phase. At very low doping of few percentage, spiral modes  promote ferromagnetism. In this paper, we showed analytically, generation of spin gaps in odd-ladders and the explicit relation between magnetic orders and topological excitation.
  In our forthcoming study, we are to examine 
 whether the spiral phase enhances superconducting transition or suppresses
  via tunneling of solitons. 
 	
\acknowledgments

This work  is supported by Creative Research Initiatives
of the Korean Ministry of Science and Technology. The author is very grateful to Prof. S.I. Lee for the valuable discussions.

\begin{figure}
 \begin{center}
 \epsfig{file=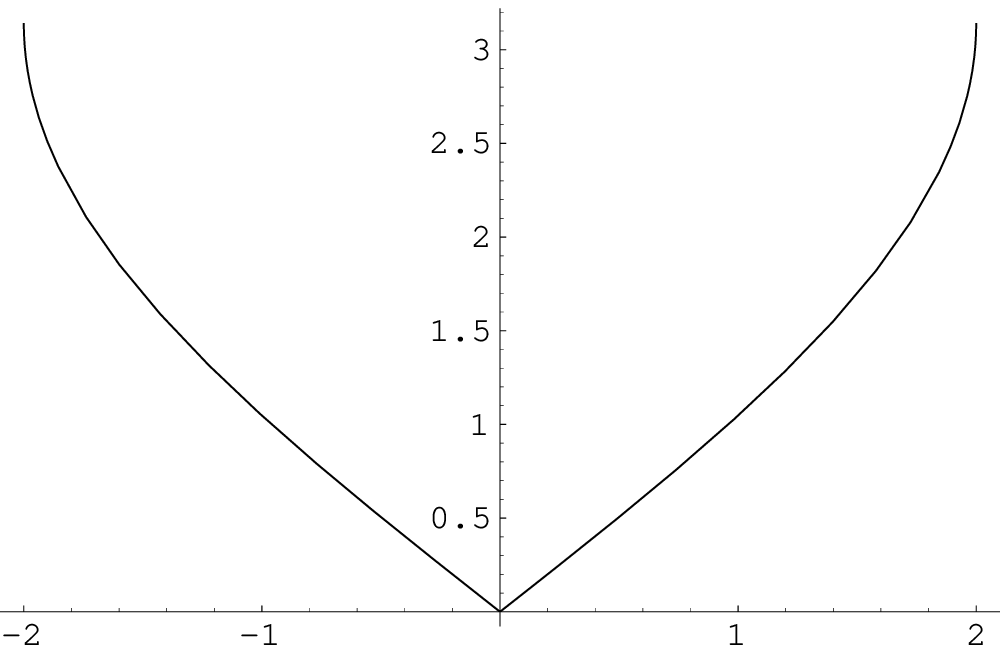,clip=,width=0.3\textwidth}
 \end{center}
 \caption{Spiral wavevector $Q_{y}$ vs. $Q_{x}$ for $0 <\phi<\pi$. 
  $Q$ reaches the maximum at $Q_{x}=0$, which indicates
  the gap formation along the $y$-direction.}
 \label{fig1} 
 \end{figure}

 \begin{figure} 
\begin{center}
    \epsfig{file=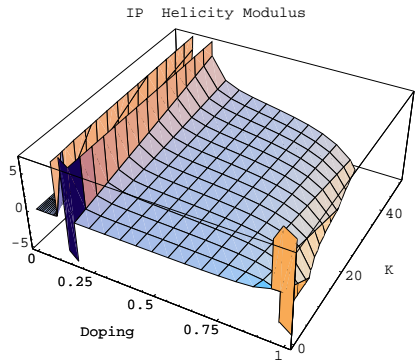,clip=,width=0.35\textwidth}
 \end{center}
\caption{IP helicity modulus(HM) for 3-leg ladder with $\widetilde{g}=1, 
 t_{\perp}=0.2$\hspace{0.1cm} eV  and $J_{\perp}=0.04$ \hspace{0.1cm}eV. The temperature is
 given in K. When HM becomes negative,  spiral phase is formed at $0.1 < \delta < 0.2$. 
 The sharp increase at higher  $\delta\ge 0.8$ with subsequent negative region indicates another window of spiral mode.  As leg-number increases, the doping regime for IP spirals is shifted to smaller values.}
 \label{fig2}
 \end{figure}
 
 \begin{figure} 
  \begin{center}
    \epsfig{file=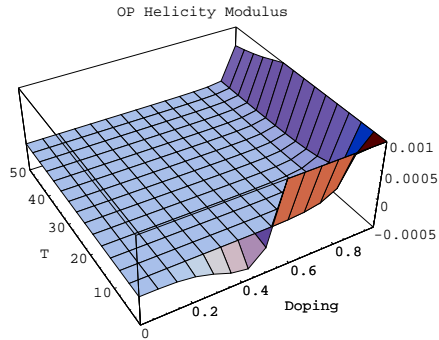,clip=,width=0.35\textwidth}
  \end{center}
 \caption{OP helicity modulus for 3-leg ladder.  OP spiral mode is formed at doping range $0<\delta \le 0.5$. Overall features are common for ladders with more number of leggs, given the weaker dependence of the HM on leg-number than the dependence on  doping or temperature.}
   \label{fig3}
  \end{figure}

\end{document}